# Relation between a force curve measured on a solvated surface and the solvation structure: Relational expressions for a binary solvent and a molecular liquid

Ken-ichi Amano[a], Kazuhiro Suzuki[b], Takeshi Fukuma[c], and Hiroshi Onishi[a]

[a]*Department of Chemistry, Faculty of Science, Kobe University, Nada-ku, Kobe 657-8501, Japan.*

[b]*Department of Electronic Science and Engineering, Kyoto University, Katsura, Nishikyo, Kyoto 615-8510, Japan.*

[c]*Bio-AFM Frontier Research Center, Kanazawa University, Kakuma-machi, Kanazawa 920-1192, Japan.*

Author to whom correspondence should be addressed: Ken-ichi Amano.

Electric mail: k-amano@gold.kobe-u.ac.jp

# 1. Introduction

Recent frequency modulated atomic force microscopy (FM-AFM) can measure force curves between a probe and a sample surface in several solvents (e.g., in a simple liquid [1], a binary solvent [2], and a molecular liquid [3-9], etc). In the present circumstances, unfortunately, the force curve is supposed to be the solvation structure in some degree, because its shape is generally oscilltive and pitch of the oscillation is about the same as diameter of the solvent particle. However, it is not the solvation structure. It is just only an interaction force between the probe and sample surface. (Hereafter, we call the interaction force as a mean force.) To elucidate the relation between the force curve and solvation structure, we have recently derived a relational expression between them for the simple liquid [10]. A usual simple liquid is composed of small hard-spheres or Lennard-Jones (LJ) spheres. The LJ spheres mean that each sphere interacts with LJ potential (e.g., Octamethylcyclotetrasiloxane (OMCTS) and carbon tetrachloride ($CCl_4$) are one of the typical LJ liquids). Although we have derived the relational expression for the simple liquid, the relational expressions for a binary solvent and a molecular liquid have not produced, yet. Hence, in the present study, we try to obtain the relational expressions within the two types of the solvents. The binary solvent denotes a mixed solvent that contains two kinds of particles. In this paper, a solvent that contains two kinds of the simple liquids or a simple ionic liquid is treated as the binary solvent. That is, the binary solvent in the present paper represents the mixed solvent that contains *two kinds of small spheres*. On the other hand, the molecular liquid focused in this paper represents, for example, water, ethanol, acetone, and so on. Since a consideration about conformational variation of a liquid molecule leads to higher complexity, an ensemble of large molecules (chain-like molecules) is not treated as the molecular liquid here. Intramolecular vibrations of stretching and bending are not under consideration, too. (These approximations are discussed in Chapter 2.A in more detail.) We derive the relational expressions in such the binary solvent and molecular liquid. The derivations are done in the base of statistical mechanics of liquid in equilibrium state [11,12].

Here, we explain theoretical conditions for the AFM in the binary solvent and molecular liquid. The probe and the sheet of the sample surface are immersed in the each solvent. The theoretical systems are treated in the canonical ensemble and the volumes of the systems are supposed to be sufficiently large. The sheet and probe immersed are treated as the solutes. Their orientations cannot be changed (are fixed), but three-dimensional placements can be changed. In the first half of the theoretical derivation, the tip apex (nano-cluster probe [13]) is treated, however, to connect the force curve and solvation structure within a simple relationship, an ideal probe is introduced. In the ideal probe, the solvent particle of the binary solvent or molecular liquid is attached to the tip apex of the probe and the other parts are neglected (i.e., the other parts do not have volumes and interactions with the solvent and the sheet). Consequently, the force curves and solvation structure are linked. The present study connects the force curve measured by the ideal probe and solvation structure through a simple equation. (Direct connection between the force curve measured by the nano-cluster probe and solvation structure is next study.)

The present study is always carried out in equilibrium state. Hence, the mean force between the sheet and probe is treated as that of conservative one, though the measured force is not exactly conservative force (however, it is almost the conservative force in some cases). In the theoretical condition, the tip apex (nano-cluster probe) and ideal probe are assumed to be the probe models. That is, only a small part of the probe is considered. This assumption is considered to be valid, because it is demonstrated by the liquid AFM [6] that the only small part of the tip apex is important for an experimental result of the force curve. In addition, since the real probe is able to measure sample surface *at molecular resolution*, it is conceivable that the only small part of its tip apex with Angstrom seize has significant role in the measurement. Therefore, we have applied only the small part in the theoretical probe model.

In this paper, we explain the relation between the force curve and solvation structure in Chapter 2 and propose the method for comparing them in Chapter 3 concerning the both binary solvent and molecular liquid. The discussions about the force curve and solvation structure are performed in a following background: The force curve is that obtained by the liquid AFM, and the solvation structure is that obtained by a calculation [5, 13-19] or an experiment [20-22]. We carry out the present study by using statistical mechanics of liquid in equilibrium state.

## 2. Theory

**A. The relational expression for a binary solvent**

Here, we introduce two kinds of particles, a and b. These particles are mixed in a system of canonical ensemble, which form a binary solvent. The numbers of the particles a and b are $N$ and $n$, respectively. A sheet of sample surface and a probe with arbitrary shape are immersed in the binary solvent. They are treated as peculiar solutes for theoretical convenience. The peculiar solutes mean that their positions can be changed within the system (artificially), but they cannot move within the system with their kinetic momenta. We construct the relational expression for the system without external field. In this case, fundamental partition function ($Q_O$) is expressed as

$$Q_O = \frac{\zeta_a^N \zeta_b^n}{N!n!} \int ... \int \exp\{-\beta U(\mathbf{r}_M', \mathbf{r}_P', \mathbf{r}_{a1}, ..., \mathbf{r}_{aN}, \mathbf{r}_{b1}, ..., \mathbf{r}_{bn})\} d\mathbf{r}_M' d\mathbf{r}_P' d\mathbf{r}_{a1}...d\mathbf{r}_{aN} d\mathbf{r}_{b1}...d\mathbf{r}_{bn}, \qquad (1)$$

where $\zeta_i$ ($i$=a or b) is

$$\zeta_i = (2\pi m_i kT)^{3/2} / h^3. \tag{2}$$

Here, **r** represents three-dimensional vector that is expressed as **r**=(*x*,*y*,*z*) or *x***i**+*y***j**+*z***k** where **i**, **j**, and **k** are unit vectors of *x*, *y*, and *z*-axes. ∫d**r** means ∭d*x*d*y*d*z*, integrations of which are performed within the system of volume *V* (i.e., volume integral). Characters of *β* and *U* represent 1 divided by "Boltzmann's constant (*k*) times temperature (*T*)" and internal energy, respectively. Subscripts M and P represent the sample surface (sheet) and probe (tip), respectively. Subscripts a*j* (*j*=1, 2, ⋯, *N*) and b*k* (*k*=1, 2, ⋯, *n*) represent particles of solvent species a and b, respectively. *π*, *m*, and *h* are circle ratio, weight of the solvent particle, and Planck's constant, respectively.

Since the mean force along *z*-axis is measured between the sample surface and probe, it is represented as $f_{\text{MP}z}$. Similarly, the mean force that includes the all components of *x*, *y*, and *z*-axes is written as **f**$_{\text{MP}}$. Pair distribution function (*g*) between the sample surface and solvent *i* (*i*=a or b) corresponds to the solvation structure, which is represented as $g_{\text{MS}i}$ where the subscript S*i* denotes the solvent species *i*. Here, the aim of this section is rewritten as "finding of the relation between the mean force and solvation structure of species *i*". The relation is symbolized to "**f**$_{\text{MP}}$↔$g_{\text{MS}i}$". Using the fundamental partition function, we explain the relation through a route, **f**$_{\text{MP}}$↔$g_{\text{MP}}$↔$g_{\text{MP}i*}$↔$g_{\text{MS}i}$, where a subscript P*i*\* represents the ideal probe for the solvent species *i* (*i*=a or b). The ideal probe means that a solvent particle *i* is jointed on the tip of the probe and the other parts of the probe are disregarded. That is, the ideal probe is composed of a solvent particle *i* and a ghost-like body (i.e., the solvent particles can go through the body). Although the ideal probe is the same as the solvent particle, the ideal probe is treated as distinguishable one from the solvent particles in the theory.

Next, we shall see a partition function with two variables, **r**$_\text{M}$ and **r**$_\text{P}$. Positions of the sheet and probe are fixed at **r**$_\text{M}$ and **r**$_\text{P}$, respectively, in the partition function (*Q*), which is expressed as

$$Q(\mathbf{r}_M,\mathbf{r}_P) = \frac{\zeta_a^N \zeta_b^n}{N!n!} \int \ldots \int \exp\{-\beta U(\mathbf{r}_M,\mathbf{r}_P,\mathbf{r}_{a1},\ldots,\mathbf{r}_{aN},\mathbf{r}_{b1},\ldots,\mathbf{r}_{bn})\} d\mathbf{r}_{a1}\ldots d\mathbf{r}_{aN} d\mathbf{r}_{b1}\ldots d\mathbf{r}_{bn}. \tag{3}$$

In Eq. (3), translational displacements of the solvent particles are performed by avoiding the sheet and probe. By the way, the mean force between the sheet and probe is written in the form

$$\mathbf{f}_{MP}(\mathbf{r}_M,\mathbf{r}_P) \equiv -\frac{\partial \Phi_{MP}(\mathbf{r}_M,\mathbf{r}_P)}{\partial \mathbf{r}_P} = -\frac{\partial \{F(\mathbf{r}_M,\mathbf{r}_P) - F(\infty,\infty)\}}{\partial \mathbf{r}_P}, \tag{4}$$

where $\Phi_{MP}$ is potential of mean force between the sheet and probe. Detail of $\Phi_{MP}$ is explained in two papers [23,24]. $F$ is free energy of the system and $(\infty,\infty)$ represents the sheet and probe are infinitely separated. Partial differentiation of vector $\mathbf{r}$ represents $\partial/\partial\mathbf{r} = (\partial/\partial x)\mathbf{i} + (\partial/\partial y)\mathbf{j} + (\partial/\partial z)\mathbf{k}$. It is defined in Eq. (4) that when a value of $\mathbf{f}_{MP}$ of $l$-component ($l=x$, $y$, or $z$) is positive the probe feels force whose direction is same as that of $l$-axis, while when the value is negative the probe feels force whose direction is opposite to $l$-axis. Since $F(\infty,\infty)$ is a constant, $\mathbf{f}_{MP}$ is rewritten as

$$\mathbf{f}_{MP}(\mathbf{r}_M,\mathbf{r}_P) = -\frac{\partial F(\mathbf{r}_M,\mathbf{r}_P)}{\partial \mathbf{r}_P} = -\frac{\partial}{\partial \mathbf{r}_P}(-kT \ln Q) = kT \frac{1}{Q}\frac{\partial Q}{\partial \mathbf{r}_P}. \tag{5}$$

On the other hand, definition of the pair distribution function between the sheet and probe is expressed as

$$g_{MP}(\mathbf{r}_M,\mathbf{r}_P) = \frac{1}{\rho_M \rho_P} \cdot \rho_{MP}(\mathbf{r}_M,\mathbf{r}_P), \tag{6}$$

where $\rho_{MP}$, $\rho_M$, and $\rho_P$ represent pair density distribution between the sheet and probe, bulk density of the sheet, and bulk density of the probe, respectively. (The two bulk densities are constants.) Detail of Eq. (6) is described as follows:

$$g_{MP}(\mathbf{r}_M, \mathbf{r}_P) = \frac{1}{\rho_M \rho_P} \cdot \left\langle \delta(\mathbf{r}_M - \mathbf{r}_M^{'}) \delta(\mathbf{r}_P - \mathbf{r}_P^{'}) \right\rangle = \frac{Q(\mathbf{r}_M, \mathbf{r}_P)}{\rho_M \rho_P Q_O},  \tag{7}$$

where <X> means the ensemble average of X. Thus, $kT(\partial/\partial \mathbf{r}_P)\ln(g_{MP})$ gives

$$kT \frac{\partial}{\partial \mathbf{r}_P} \ln g_{MP}(\mathbf{r}_M, \mathbf{r}_P) = kT \frac{1}{Q} \frac{\partial Q}{\partial \mathbf{r}_P}. \tag{8}$$

Since right-hand sides of Eqs. (5) and (8) are the same, the relational expression for $\mathbf{f}_{MP}$ and $g_{MP}$ ($f_{MP} \leftrightarrow g_{MP}$) can be written as

$$\mathbf{f}_{MP}(\mathbf{r}_M, \mathbf{r}_P) = kT \frac{\partial}{\partial \mathbf{r}_P} \ln g_{MP}(\mathbf{r}_M, \mathbf{r}_P). \tag{9}$$

Additionally, when the probe is changed to the ideal one, following transformation can readily be done:

$$kT \frac{\partial}{\partial \mathbf{r}_P} \ln g_{MP}(\mathbf{r}_M, \mathbf{r}_P) \bigg|_{P \to Pi^*} = kT \frac{\partial}{\partial \mathbf{r}_{Pi^*}} \ln g_{MPi^*}(\mathbf{r}_M, \mathbf{r}_{Pi^*}), \qquad \text{where } i=\text{a or b}. \tag{10}$$

Eq. (10) gives relation between $g_{MP}$ and $g_{MPi^*}$ ($g_{MP} \leftrightarrow g_{MPi^*}$). The probe with arbitrary shape is changed to the ideal probe in Eq. (10). Although the change is difficult in a real AFM system, the change can readily be done in the theoretical system, because the shape of the probe is not specified in the function of $U$. To achieve the goal of this section, we shall compare $g_{MPi^*}$ and $g_{MSi}$ next. Since $\rho_M$ and $\rho_{Pi^*}$ are both $(1/V)$, $g_{MPi^*}$ is calculated to be

$$g_{MPi^*}(\mathbf{r}_M, \mathbf{r}_{Pi^*}) = \frac{V^2 \int..\int \exp\{-\beta U(\mathbf{r}_M, \mathbf{r}_{Pi^*}, \mathbf{r}_{a1},...,\mathbf{r}_{aN}, \mathbf{r}_{b1},...,\mathbf{r}_{bn})\} d\mathbf{r}_{a1}..d\mathbf{r}_{aN} d\mathbf{r}_{b1}..d\mathbf{r}_{bn}}{\int..\int \exp\{-\beta U(\mathbf{r}_M^{'}, \mathbf{r}_{Pi^*}^{'}, \mathbf{r}_{a1},...,\mathbf{r}_{aN}, \mathbf{r}_{b1},...,\mathbf{r}_{bn})\} d\mathbf{r}_M^{'} d\mathbf{r}_{Pi^*}^{'} d\mathbf{r}_{a1}..d\mathbf{r}_{aN} d\mathbf{r}_{b1}..d\mathbf{r}_{bn}}, \tag{11}$$

where $V$ is volume of the system. If the ideal probe is located at $\mathbf{r}_{Si}$ ($\mathbf{r}_{Pi*} \to \mathbf{r}_{Si}$) and $\mathbf{r}_{Pi*}'$ at denominator are alternated to $\mathbf{r}_{i0}$ ($\mathbf{r}_{Pi*}' \to \mathbf{r}_{i0}$), Eq. (11) is rewritten as

$$g_{MPi*}(\mathbf{r}_M, \mathbf{r}_{Si}) = \frac{V^2 \int..\int \exp\{-\beta U(\mathbf{r}_M, \mathbf{r}_{Si}, \mathbf{r}_{a1},...,\mathbf{r}_{aN}, \mathbf{r}_{b1},...,\mathbf{r}_{bn})\} d\mathbf{r}_{a1}..d\mathbf{r}_{aN} d\mathbf{r}_{b1}..d\mathbf{r}_{bn}}{\int..\int \exp\{-\beta U(\mathbf{r}_M', \mathbf{r}_{i0}, \mathbf{r}_{a1},...,\mathbf{r}_{aN}, \mathbf{r}_{b1},...,\mathbf{r}_{bn})\} d\mathbf{r}_M' d\mathbf{r}_{i0} d\mathbf{r}_{a1}..d\mathbf{r}_{aN} d\mathbf{r}_{b1}..d\mathbf{r}_{bn}}, \quad (12)$$

Furthermore, setting $i=a$ (considering the ideal probe of the solvent species a), Eq. (12) is given by

$$g_{MPa*}(\mathbf{r}_M, \mathbf{r}_{Sa}) = \frac{V^2 \int...\int \exp\{-\beta U_A(\mathbf{r}_M, \mathbf{r}_{Sa}, \mathbf{r}_{a2},...,\mathbf{r}_{aN+1}, \mathbf{r}_{b1},...,\mathbf{r}_{bn})\} d\mathbf{r}_{a2}...d\mathbf{r}_{aN+1} d\mathbf{r}_{b1}...d\mathbf{r}_{bn}}{\int...\int \exp\{-\beta U_A(\mathbf{r}_M', \mathbf{r}_{a1},...,\mathbf{r}_{aN+1}, \mathbf{r}_{b1},...,\mathbf{r}_{bn})\} d\mathbf{r}_M' d\mathbf{r}_{a1}...d\mathbf{r}_{aN+1} d\mathbf{r}_{b1}...d\mathbf{r}_{bn}}, \quad (13)$$

where subscript A adjacent to $U$ means "absence of the probe" (i.e., $U_A$ does not include the probe in the internal energy). On the other hand, the $g_{MSi}$ is expresses as

$$g_{MSi}(\mathbf{r}_M, \mathbf{r}_{Si}) = \frac{\rho_{MSi}(\mathbf{r}_M, \mathbf{r}_{Si})}{\rho_M \rho_{Si}} = \frac{1}{\rho_M \rho_{Si}} \cdot \left\langle \delta(\mathbf{r}_M - \mathbf{r}_M') \sum_{j=1}^{c} \delta(\mathbf{r}_{Si} - \mathbf{r}_{ij}) \right\rangle, \quad \text{where } i=\text{a or b.} \quad (14)$$

In Eq. (14), the ensemble average must be calculated in the absence of the probe, because $g_{MSi}$ corresponds to the pure solvation structure formed on the sheet. A small letter "c" on the summation ($\Sigma$) takes $N$ when $i=$a and $n$ when $i=$b. If the probe exists, the binary solvent is sandwiched between the sheet and probe, which brings about destruction of the pure solvation structure. Thus, the ensemble average must be calculated with $U_A$. If $i=$a, Eq. (14) is calculated to be

$$g_{MSa}(\mathbf{r}_M, \mathbf{r}_{Sa}) = \frac{V^2 \int...\int \exp\{-\beta U_A(\mathbf{r}_M, \mathbf{r}_{Sa}, \mathbf{r}_{a2},...,\mathbf{r}_{aN}, \mathbf{r}_{b1},...,\mathbf{r}_{bn})\} d\mathbf{r}_{a2}...d\mathbf{r}_{aN} d\mathbf{r}_{b1}..d\mathbf{r}_{bn}}{\int...\int \exp\{-\beta U_A(\mathbf{r}_M', \mathbf{r}_{a1},...,\mathbf{r}_{aN}, \mathbf{r}_{b1},...,\mathbf{r}_{bn})\} d\mathbf{r}_M' d\mathbf{r}_{a1}...d\mathbf{r}_{aN} d\mathbf{r}_{b1}...d\mathbf{r}_{bn}}, \quad (15)$$

where we used $\rho_M=1/V$ and $\rho_{Sa}=N/V$. Then, comparing Eqs. (13) and (15), it is revealed that $g_{MPa*}$ is (fairly) equal to $g_{MSa}$ when $N$ is sufficiently large ($1<<N$). Also, this conclusion consists in the case of $i$=b (and $1<<n$). By the way, if an infinite number of the solvent particles exist in the system with constant volume $V$, the internal energy becomes infinite due to extremely high crowding of the solvent particles, which implies that the system cannot be exist completely. That is, when $N$ ($n$) is infinite, the system loses the physical meaning (i.e., the ensemble is neither fluid nor solid). Therefore, the range of $N$ ($n$) is considered to be $1<<N<\infty$ ($1<<n<\infty$).

By summing up above results, the relational expression ($\mathbf{f}_{MP} \leftrightarrow g_{MSi}$) for the binary solvent is given by

$$\mathbf{f}_{MP}(\mathbf{r}_M, \mathbf{r}_P)\big|_{P \to Pi*} = \mathbf{f}_{MPi*}(\mathbf{r}_M, \mathbf{r}_{Pi*})\big|_{\mathbf{r}_{Pi*} \to \mathbf{r}_{Si}} = \mathbf{f}_{MPi*}(\mathbf{r}_M, \mathbf{r}_{Si}) = kT \frac{\partial}{\partial \mathbf{r}_{Si}} \ln g_{MSi}(\mathbf{r}_M, \mathbf{r}_{Si}), \quad (16)$$

where $i$=a or b and $\mathbf{r}_{Pi*} \to \mathbf{r}_{Si}$ means that only the character is replaced. That is, its vector value is not changed. It is realized that the relational expression for the binary solvent (Eq. (16)) is very similar to that for the simple liquid [25].

### B. The relational expression for a molecular liquid

In this section, we derive the relational expression for a molecular liquid. As explained in Chapter 1, the molecular liquid represents water, ethanol, acetone, etc. Since consideration about conformational variation of a liquid molecule leads to higher complexity of the theory, a large molecule (e.g., a chain-like molecule) is not treated as the liquid molecule (i.e., a molecule that composed of two, three, or several atoms is treated as the liquid molecule). Here, we introduce two kinds of the liquid molecules.

One is a linear molecule, the other is a non-linear molecule. The linear molecule represents a diatomic molecule or a rod-like molecule (e.g., $N_2$, $O_2$, $C_2H_2$), and the non-linear molecule represents a non-rod-like molecule (e.g., $H_2O$, $CH_3OH$, $CH_3COCH_3$). We discuss the molecular liquid in a system of canonical ensemble, where the number of the liquid molecules is $N$. A sheet of sample surface and a probe with arbitrary shapes are immersed in the molecular liquid. They are treated as peculiar solutes for theoretical convenience. The peculiar solutes mean that their positions can be changed within the system (artificially), but they cannot move within the system with their kinetic momenta. A theoretically different point against the probes for the simple and binary solvents is that the probe for the molecular liquid is set to have orientational degree of freedom ($\mathbf{\Omega}_P$: $\mathbf{\Omega}$ represents Euler angle) from the start point of the theory. Importance of the setting is realized in the course of the derivation. Considering the system without external field, fundamental partition function ($Q_O$) is expressed as

$$Q_O = \frac{\zeta_{all}^N}{N!} \int..\int \exp\{-\beta U(\mathbf{r}_M', \mathbf{r}_P', \mathbf{r}_1,...,\mathbf{r}_N, \mathbf{\Omega}_P', \mathbf{\Omega}_1,...,\mathbf{\Omega}_N)\} d\mathbf{r}_M' d\mathbf{r}_P' d\mathbf{r}_1..d\mathbf{r}_N d\mathbf{\Omega}_P' d\mathbf{\Omega}_1..d\mathbf{\Omega}_N . \quad (17)$$

Here, $\zeta_{all}$ times $h^3$ is a constant that contains all of the integrated values of kinetic momenta of the liquid molecule. In the construction of $Q_O$, conformational variation of the liquid molecule and intramolecular vibrations of stretching and bending are neglected. This is because, the liquid molecule considered here is a smaller molecule whose basic structure is fixed in one structure. These hypotheses (approximations) are acceptable when (potential of) the mean force and solvation structure are studied in its normal liquid phase. (If an ensemble of chain-like molecules is supposed to be the molecular liquid, in a strict sense, the conformational variation must be incorporated in its partition function. Furthermore, if *absolute* heat capacity of the molecular liquid is the topic of this study, the intramolecular vibrations of stretching and bending must be treated within a theory of quantum statistical mechanics. However, these complexities are able to be ignored in this study.) Here, we rewrite the aim of this section as follows: "finding of the relational expression between the mean force and solvation structure on the sample surface for the molecular liquid". The relation is symbolized to "$\mathbf{f}_{MP} \leftrightarrow g_{MS}$". Using the fundamental partition function, we explain the relation through a route, $\mathbf{f}_{MP} \leftrightarrow g_{MP} \leftrightarrow g_{MP*} \leftrightarrow g_{MS}$, where the subscript P* represents the ideal probe for the molecular liquid. The ideal probe means that the liquid molecule is jointed on the tip of the probe with Euler angle $\mathbf{\Omega}_{P*}$ and the other parts of the probe are disregarded. That is, the ideal probe is composed of the liquid molecule and a ghost-like body.

Next, we shall see a partition function with three variables, $\mathbf{r}_M$, $\mathbf{r}_P$, and $\mathbf{\Omega}_P$. Positions of the sheet and probe are fixed at $\mathbf{r}_M$ and $\mathbf{r}_P$, respectively, and the orientation of the probe is fixed at $\mathbf{\Omega}_P$ in the partition function ($Q$), which is expressed as

$$Q(\mathbf{r}_M, \mathbf{r}_P, \mathbf{\Omega}_P) = \frac{\zeta_{\text{all}}^N}{N!} \int ..\int \exp\{-\beta U(\mathbf{r}_M, \mathbf{r}_P, \mathbf{r}_1, ..., \mathbf{r}_N, \mathbf{\Omega}_P, \mathbf{\Omega}_1, ..., \mathbf{\Omega}_N)\} d\mathbf{r}_1 .. d\mathbf{r}_N d\mathbf{\Omega}_1 .. d\mathbf{\Omega}_N . \qquad (18)$$

In Eq. (18), translational and orientational displacements of the liquid molecules are performed by avoiding the sheet and probe. By the way, we shall see the mean force between the sheet and probe in the molecular liquid here, which is expressed as (refer Eqs. (4) and (5))

$$\mathbf{f}_{MP}(\mathbf{r}_M, \mathbf{r}_P, \mathbf{\Omega}_P) = kT \frac{1}{Q} \frac{\partial Q}{\partial \mathbf{r}_P} . \qquad (19)$$

On the other hand, the pair distribution function between the sheet and probe is written as

$$g_{MP}(\mathbf{r}_M, \mathbf{r}_P, \mathbf{\Omega}_P) = \frac{\omega_P}{\rho_M \rho_P} \cdot \rho_{MP}(\mathbf{r}_M, \mathbf{r}_P, \mathbf{\Omega}_P) , \qquad (20)$$

where $\omega_P$ is an integrated value of Euler angle of the probe. When shape of the probe is linear, $\omega_P$ is written in the form

$$\omega_P \equiv \int_0^{2\pi} \int_0^{\pi} \sin\theta d\theta d\phi = 4\pi, \qquad (21)$$

while when the shape is non-linear, $\omega_P$ is given by

$$\omega_P \equiv \int_0^{2\pi} \int_0^{2\pi} \int_0^{\pi} \sin\theta d\theta d\phi d\chi = 8\pi^2. \qquad (22)$$

Above definitions are also applied for that of the ideal probe ($\omega_{P*}$) and liquid molecule ($\omega_S$). Then, Eq. (20) is rewritten as

$$g_{MP}(\mathbf{r}_M, \mathbf{r}_P, \mathbf{\Omega}_P) = \frac{\omega_P}{\rho_M \rho_P} \cdot \left\langle \delta(\mathbf{r}_M - \mathbf{r}_M')\delta(\mathbf{r}_P - \mathbf{r}_P')\delta(\mathbf{\Omega}_P - \mathbf{\Omega}_P') \right\rangle = \frac{\omega_P}{\rho_M \rho_P Q_O} Q(\mathbf{r}_M, \mathbf{r}_P, \mathbf{\Omega}_P), \qquad (23)$$

Thus, $kT(\partial/\partial \mathbf{r}_P)\ln(g_{MP})$ is calculated to be

$$kT \frac{\partial}{\partial \mathbf{r}_P} \ln g_{MP}(\mathbf{r}_M, \mathbf{r}_P, \mathbf{\Omega}_P) = kT \frac{1}{Q} \frac{\partial Q}{\partial \mathbf{r}_P}. \qquad (24)$$

Since right-hand sides of Eqs. (19) and (24) are the same, the relational expression for $\mathbf{f}_{MP}$ and $g_{MP}$ ($f_{MP} \leftrightarrow g_{MP}$) is

$$\mathbf{f}_{MP}(\mathbf{r}_M, \mathbf{r}_P, \mathbf{\Omega}_P) = kT \frac{\partial}{\partial \mathbf{r}_P} \ln g_{MP}(\mathbf{r}_M, \mathbf{r}_P, \mathbf{\Omega}_P). \qquad (25)$$

Additionally, when the probe is changed to the ideal one, following transformation can readily be done:

$$kT\frac{\partial}{\partial \mathbf{r}_P}\ln g_{MP}(\mathbf{r}_M,\mathbf{r}_P,\mathbf{\Omega}_P)\bigg|_{P\to P*} = kT\frac{\partial}{\partial \mathbf{r}_{P*}}\ln g_{MP*}(\mathbf{r}_M,\mathbf{r}_{P*},\mathbf{\Omega}_{P*}). \qquad (26)$$

Eq. (26) gives relation between $g_{MP}$ and $g_{MP*}$ ($g_{MP}\leftrightarrow g_{MP*}$). To achieve the goal of this section, we shall compare $g_{MP*}$ and $g_{MS}$ next. Since $\rho_M$ and $\rho_{P*}$ are both $(1/V)$, $g_{MP*}$ is calculated to be

$$g_{MP*}(\mathbf{r}_M,\mathbf{r}_{P*},\mathbf{\Omega}_{P*}) = \frac{V^2 \omega_{P*}\int..\int \exp\{-\beta U(\mathbf{r}_M,\mathbf{r}_{P*},\mathbf{r}_1,..,\mathbf{r}_N,\mathbf{\Omega}_{P*},\mathbf{\Omega}_1,..,\mathbf{\Omega}_N)\}d\mathbf{r}_1..d\mathbf{r}_N d\mathbf{\Omega}_1..d\mathbf{\Omega}_N}{\int..\int \exp\{-\beta U(\mathbf{r}_M',\mathbf{r}_{P*}',\mathbf{r}_1,..,\mathbf{r}_N,\mathbf{\Omega}_{P*}',\mathbf{\Omega}_1,..,\mathbf{\Omega}_N)\}d\mathbf{r}_M'd\mathbf{r}_{P*}'d\mathbf{r}_1..d\mathbf{r}_N d\mathbf{\Omega}_{P*}'d\mathbf{\Omega}_1..d\mathbf{\Omega}_N}. \qquad (27)$$

Using an equation $\omega_{P*}=\omega_S$ and referring the method to derive Eq. (13) from Eq. (11), $g_{MP*}$ is rewritten as

$$g_{MP*}(\mathbf{r}_M,\mathbf{r}_S,\mathbf{\Omega}_S) = \frac{V^2 \omega_S \int..\int \exp\{-\beta U_A(\mathbf{r}_M,\mathbf{r}_S,\mathbf{r}_2,..,\mathbf{r}_{N+1},\mathbf{\Omega}_S,\mathbf{\Omega}_2,..,\mathbf{\Omega}_{N+1})\}d\mathbf{r}_2..d\mathbf{r}_{N+1}d\mathbf{\Omega}_2..d\mathbf{\Omega}_{N+1}}{\int..\int \exp\{-\beta U_A(\mathbf{r}_M',\mathbf{r}_1,..,\mathbf{r}_{N+1},\mathbf{\Omega}_1,..,\mathbf{\Omega}_{N+1})\}d\mathbf{r}_M'd\mathbf{r}_1..d\mathbf{r}_{N+1}d\mathbf{\Omega}_1..d\mathbf{\Omega}_{N+1}}. \qquad (28)$$

On the other hand, the pair distribution function between the sheet and the ensemble of liquid molecules ($g_{MS}$) is expressed as

$$g_{MS}(\mathbf{r}_M,\mathbf{r}_S,\mathbf{\Omega}_S) = \frac{\omega_S \rho_{MS}(\mathbf{r}_M,\mathbf{r}_S,\mathbf{\Omega}_S)}{\rho_M \rho_S} = \frac{\omega_S}{\rho_M \rho_S}\cdot\left\langle \delta(\mathbf{r}_M-\mathbf{r}_M')\sum_{i=1}^{N}\delta(\mathbf{r}_S-\mathbf{r}_i)\delta(\mathbf{\Omega}_S-\mathbf{\Omega}_i)\right\rangle. \qquad (29)$$

Using $\rho_M=1/V$ and $\rho_S=N/V$, the $g_{MS}$ is calculated to be

$$g_{MS}(\mathbf{r}_M,\mathbf{r}_S,\mathbf{\Omega}_S) = \frac{V^2 \omega_S \int..\int \exp\{-\beta U_A(\mathbf{r}_M,\mathbf{r}_S,\mathbf{r}_2,..,\mathbf{r}_N,\mathbf{\Omega}_S,\mathbf{\Omega}_2,..,\mathbf{\Omega}_N)\}d\mathbf{r}_2..d\mathbf{r}_N d\mathbf{\Omega}_2..d\mathbf{\Omega}_N}{\int..\int \exp\{-\beta U_A(\mathbf{r}_M',\mathbf{r}_1,..,\mathbf{r}_N,\mathbf{\Omega}_1,..,\mathbf{\Omega}_N)\}d\mathbf{r}_M'd\mathbf{r}_1..d\mathbf{r}_N d\mathbf{\Omega}_1..d\mathbf{\Omega}_N}. \qquad (30)$$

Then, comparing Eqs. (28) and (30), it is revealed that $g_{MP*}$ is (fairly) equal to $g_{MS}$ when $N$ is sufficiently large ($1 \ll N$). As explained in the previous section (Chapter 2.A), the range of $N$ that still possessing the physical meaning is considered to be $1 \ll N < \infty$ (i.e., the system with constant volume cannot contain an infinite number of the liquid molecules).

By summing up above results, the relational expression ($\mathbf{f}_{MP} \leftrightarrow g_{MS}$) for the molecular liquid is given by

$$\mathbf{f}_{MP}(\mathbf{r}_M, \mathbf{r}_P, \mathbf{\Omega}_P)\big|_{P \to P*}$$
$$= \mathbf{f}_{MP*}(\mathbf{r}_M, \mathbf{r}_{P*}, \mathbf{\Omega}_{P*})\big|_{\mathbf{r}_{P*} \to \mathbf{r}_S, \mathbf{\Omega}_{P*} \to \mathbf{\Omega}_S} = \mathbf{f}_{MP*}(\mathbf{r}_M, \mathbf{r}_S, \mathbf{\Omega}_S) = kT \frac{\partial}{\partial \mathbf{r}_S} \ln g_{MS}(\mathbf{r}_M, \mathbf{r}_S, \mathbf{\Omega}_S) \quad (31)$$

Here, $\mathbf{r}_{P*} \to \mathbf{r}_S$ and $\mathbf{\Omega}_{P*} \to \mathbf{\Omega}_S$ mean that only the characters are replaced. That is, values of the vector and Euler angle are not changed. It is realized that the relational expression for the molecular liquid is also very similar to that for the simple liquid [25] and the binary solvent (see Eq. (16)).

## 3. Discussion

The relational expressions for the binary solvent and molecular liquid connect the mean force measured by the liquid AFM and the solvation structure obtained by a calculation or an experiment. It is not until the detailed routes for the connections are realized, the mean force and solvation structure are rightly compared. In the real AFM system, in general, the probe always exists on the upper side of the sample surface and measured mean forces are that (almost) along $z$-axis. Therefore, we include the above two general things in discussion about the method for comparison between the mean force and solvation structure. That is, following settings are included in the discussion: $z_M<z_P$, $z_M<z_{Pi*}$, and $f_{MPiz}=\mathbf{f}_{MPi}\cdot\mathbf{k}$ ($i$=a or b) for the binary solvent; $z_M<z_P$, $z_M<z_{P*}$, and $f_{MPz}=\mathbf{f}_{MP}\cdot\mathbf{k}$ for the molecular liquid. The procedures for the comparisons between the mean forces and solvation structures are as follows:

(I) Measure $f_{MPz}$ by using liquid AFM.

(II) Obtain $g_{MSi}$ ($g_{MS}$) from a calculation or an experiment.

(III) Calculate $f_{MPi*z}$ ($f_{MP*z}$) by substituting the $g_{MSi}$ ($g_{MS}$) into Eq. (32) or (33) below. Eqs. (32) and (33) are that for the binary solvent and molecular liquid, respectively. It is hypothecated that the ideal probe exists in the system of (II) (not in the real AFM system of (I)).

(IV) Compare shapes of the $f_{MPz}$ and $f_{MPi*z}$ ($f_{MP*z}$). When the $f_{MPz}$ is well accorded with the $f_{MPi*z}$ ($f_{MP*z}$), the probe used in the real AFM system is considered to be an almost the ideal probe. In this case, solvation structure can approximately be estimated from the $f_{MPz}$ using Eq. (34) or (35) below. Eqs. (34) and (35) are that for the binary solvent and molecular liquid, respectively. On the other hand, when the $f_{MPz}$ is not similar to the $f_{MPi*z}$ ($f_{MP*z}$), it is exposed that the probe used in the AFM system is clearly different from the ideal probe.

$$kT\frac{\partial}{\partial z_{Si}}\ln g_{MSi}(\mathbf{r}_M,\mathbf{r}_{Si})\bigg|_{\mathbf{r}_{Si}\to\mathbf{r}_{Pi*}}=f_{MPi*z}(\mathbf{r}_M,\mathbf{r}_{Pi*}), \qquad \text{where } i\text{=a or b.} \qquad (32)$$

$$kT\frac{\partial}{\partial z_S}\ln g_{MS}(\mathbf{r}_M,\mathbf{r}_S,\mathbf{\Omega}_S)\bigg|_{\mathbf{r}_S\to\mathbf{r}_{P*},\mathbf{\Omega}_S\to\mathbf{\Omega}_{P*}}=f_{MP*z}(\mathbf{r}_M,\mathbf{r}_{P*},\mathbf{\Omega}_{P*}). \qquad (33)$$

$$\exp\left\{-\beta\int_{z_P}^{\infty}f_{MPz}(\mathbf{r}_M,x_P,y_P,z_P')dz_P'\right\}\bigg|_{\mathbf{r}_P\to\mathbf{r}_{Si}}\approx g_{MSi}(\mathbf{r}_M,\mathbf{r}_{Si}), \qquad \text{where } i\text{=a or b.} \qquad (34)$$

$$\exp\left\{-\beta\int_{z_P}^{\infty} f_{MPz}(\mathbf{r}_M, x_P, y_P, z_P{}', \mathbf{\Omega}_P)dz_P{}'\right\}\bigg|_{\mathbf{r}_P\to\mathbf{r}_S, \mathbf{\Omega}_P\to\mathbf{\Omega}_S} \approx g_{MS}(\mathbf{r}_M, \mathbf{r}_S, \mathbf{\Omega}_S). \tag{35}$$

As mentioned before, replacements of $\mathbf{r}_S \to \mathbf{r}_{P*}$ and $\mathbf{\Omega}_S \to \mathbf{\Omega}_{P*}$, and so on, mean that only the characters are replaced. That is, values of the vectors and Euler angles are not changed. If the probe used in the AFM measurement is solvophilic (or neutral), the probe is thought to be rather similar to the ideal probe compared to the solvophobic probe. Therefore, the solvophilic (or neutral) probe should be used in order to obtain the (approximated) solvation structure from Eq. (34) or (35).

## 4. Conclusions

In the present article, we have shown how the mean force measured by liquid AFM and the solvation structure obtained by a calculation or an experiment can be connected in the binary solvent and molecular liquid. We have explained a method for comparing the mean force and solvation structure. An interesting point of this study is that if the ideal probe is used in the measurement, the solvation structure can be obtained through Eq. (34) or (35), although the existence of the ideal itself deforms the solvation structure on the sample surface.

The relational expression between $\mathbf{f}_{MP}$ and $g_{MSi}$ ($g_{MS}$) has been derived through the route: $f_{MP}\leftrightarrow g_{MP}\leftrightarrow g_{MPi*}(g_{MP*})\leftrightarrow g_{MSi}(g_{MS})$. As described in the recent our paper [25], if a hypothesis of "$\Phi_{MPi*}=\Phi_{MSi}$ ($\Phi_{MP*}=\Phi_{MS}$)" is employed, the relational expression is derived through a route: $f_{MP}\leftrightarrow g_{MP}\leftrightarrow g_{MPi*}(g_{MP*})\leftrightarrow \Phi_{MPi*}(\Phi_{MP*})\leftrightarrow \Phi_{MSi}(\Phi_{MS})\leftrightarrow g_{MSi}(g_{MS})$. Although the latter route is not shown here, the latter is easier to derive in comparison with the former, because there is the introduction of the hypothesis. In this paper, we have shown only the former route. This is because, the former is considered to be a strict route compared to the latter.

The introduction of the ideal probe has been readily performed in the derivation. This is because, in the theory of the first half, the shape of the probe is not restricted and it can take arbitrary shape. In this case, the change from the probe with arbitrary shape to the ideal probe can readily be done. Hence, $g_{MP}$ and $g_{MP*}$ are immediately connected in the derivation.

In the real AFM experiment, most of the probes are not identical one. This fact requires another method in comparison between the force curve and solvation structure. The alternative method is transformation of the measured force curve into the solvation structure, and the transformed solvation structure is compared with the solvation structure obtained by a calculation or an experiment. Recently, K. Amano [26] has proposed the method for calculating solvation structure from the measured force curve within one-dimensional model system. In the method, a sufficiently large sphere is modeled as the sample surface and a sphere with certain diameter is modeled as the probe. The transformation can be done even when the probe is either (highly) solvophilic or solvophobic, which is a different point against Eqs. (34) and (35). However, there are problems in the transformation method. The method is restricted in the one-dimensional model system and shapes of the models of the sample surface and probe are fixed in spherical shapes. Solving of the problems and development of the transformation method into the three-dimensional model system are our next challenges.

If it had been known in the first place that the probe used in the real AFM system was almost the ideal probe, the solvation structure can be estimated from the force curve through Eq. (34) or (35). It implies that development of a nano-technology which can fabricates the probe with almost ideal one is a key technology for obtaining the solvation structure from the liquid AFM. Hence, we remark that it should be studied beforehand by a simulation that what kind of the probe is the most identical probe within the commercially available probes. We believe *such kinds of studies* provide significant information for fabrication of the nearly identical probe.

In the near future, it is likely that the time for a simulation of the mean force between the sample surface and *the probe with arbitrary shape* is shortened much. It enables us to compare the measured and simulated force curves easier. However, this comparison does not provide the information about the solvation structure purely formed on the sample surface whose structure is not sandwiched between the surfaces of the sample and probe. To extract the information about the solvation structure from the measurement, it is imperative to *theoretically* capture the relation between them. Therefore, we have derived the relational expressions between them. We believe that this study deepens understanding of the mean force measured by the AFM in liquid and sheds light on the measurement of the solvation structure on the sample surface.

**Acknowledgements**

We appreciate Masahiro Kinoshita (Kyoto University) and Hiraku Oshima (Kyoto University) for helpful advices. We thank Kei Kobayashi (Kyoto University) and Shinichiro Ido (Kyoto University) for useful comments. This work was supported by Grant-in-Aid for JSPS (Japan Society for the Promotion of Science) fellows and Foundation of Advanced Technology Institute.

1st Submission: Wed., 26 Dec. 2012 03:51:57 EST.

2nd Submission: Mon., 31 Dec. 2012 EST.

(The main text was modified.)

3rd Submission: Fri., 1 Feb. 2013 EST.

(Coauthors and additional acknowledgements are added.)

4th Submission: Wed., 6 Mar. 2013 EST.

(The main text and acknowledgements are modified.)